\documentclass{article}[15pt]

\usepackage{epsfig,rotating}

\usepackage[sort&compress]{natbib}
\bibpunct{(}{)}{,}{n}{}{,}

\begin{document}

\title{
Diminishing Return for Increased Mappability with Longer Sequencing Reads:
Implications of the $k$-mer Distributions in the Human Genome
\vspace{0.2in}
\author{
Wentian Li$^{\rm 1}$\footnote{To whom correspondence should be addressed.  Email: wtli2012@gmail.com},
Jan Freudenberg$^{\rm 1}$,
Pedro Miramontes$^2$ \\
{\small \sl 1. The Robert S. Boas Center for Genomics and Human Genetics}\\
{\small \sl  The Feinstein Institute for Medical Research}\\
{\small \sl  North Shore LIJ Health System}\\
{\small \sl Manhasset, 350 Community Drive, NY 11030, USA.}\\
{\small \sl 2.  Departamento de Matem\'{a}ticas, Facultad de Ciencias}\\
{\small \sl Universidad Nacional Aut\'{o}noma de M\'{e}xico, Circuito Exterior}\\
{\small \sl Ciudad Universitaria,  M\'{e}xico 04510 DF, M\'{e}xico.  }\\
}
\date{}
}
\maketitle  
\markboth{\sl Li, Freudenberg, Miramontes) }{\sl Li, Freudenberg, Miramontes}

\large

\newpage

\begin{abstract}
The amount of non-unique sequence (non-singletons) in a genome directly 
affects the difficulty of read alignment to a reference assembly for 
high throughput-sequencing data.  Although a greater length increases the chance
for reads being uniquely mapped to the reference genome, a quantitative 
analysis of the influence of read lengths on mappability has been 
lacking. To address this question, we evaluate the $k$-mer distribution 
of the human reference genome.  The $k$-mer frequency is determined 
for $k$ ranging from 20 to 1000 basepairs.  We use the proportion of 
non-singleton $k$-mers to evaluate the mappability of reads for a 
corresponding read length.  We observe that the proportion of 
non-singletons decreases slowly with increasing $k$, and can be 
fitted by piecewise power-law functions with different exponents 
at different $k$ ranges. A faster decay at smaller values for $k$ indicates 
more limited gains for read lengths $>$ 200 basepairs.  
The frequency distributions of $k$-mers exhibit long tails in a power-law-like trend, 
and rank frequency plots exhibit a concave Zipf's curve.  The location 
of the most frequent 1000-mers comprises 172 kilobase-ranged regions, 
including four large stretches on chromosomes 1 and X, containing
genes with biomedical implications.  
Even the read length 1000 would be
insufficient to reliably sequence these specific regions.
\end{abstract}

\newpage
\section*{Introduction}

Many applications of next-generation-sequencing (NGS) in human genetic
and medical studies  depend on the ability to uniquely align DNA reads
to the human reference genome \citep{rozowsky,cahill,koehler,derrien,lee,storvall}.
This in turn is related to the level of redundancy caused by repetitive sequences in
the human genome, well known in the earlier human whole-genome shotgun
sequencing  \cite{myers,green,scherer}, and the read length $k$. When the read length is too
short, it is theoretically impossible to construct a sequence with
size comparable to the human genome that does not contain any repeats
of windows with $k$ bases.  It has been shown using graph theory that the
shortest DNA sequences avoiding any repeats of $k$-mers can be
constructed by packing all unique $k$-mers shifting one position at the
time \citep{fraenkel}.  The number of different $k$-mer types is
$4^k/2$ ($k$ even) or $(4^k + 2^k)/2$ ($k$ odd) if both a subsequence
and its reverse complement are considered to belong to the same $k$-mer type.
Solving $4^k/2 \approx 3 \times 10^9$ leads to the conclusion that
read length $k$ must be at least greater than 17 for all reads to be uniquely alignable to a
hypothetical reference sequence that has the size of the human genome.

However, in reality the human genome did not evolve by a first
principle to be consistently compact and incompressible.
Redundant sequences in the human genome have resulted from
duplication, insertion of transposable elements,
and tandem repeats due to replication slippage,
and more than half of the human genome can be traced to repetitive
transposable elements.  Although locally duplicated sequences can
be deleterious \citep{stoppa} or disease-causing \cite{conrad}, a certain level of redundancy
is crucial for biological novelty and adaptation \citep{ohno,nowak,fortna}.
For higher eukaryotes,  a slower removal of the deleterious
repeats due to low mutation rates and smaller population sizes
\citep{plotkin} lead to a higher level of genome-wide redundancy.
This in turns may lead to more protein sequences with internal
repeats and perhaps new fold or new functions such as the
case for connection tissue, cytoskeletal, and muscle proteins \citep{marcotte}.

Therefore,  $k$=17 is a very unrealistic estimation of the minimal
read length required for a perfectly successful NGS reads alignment.
Accordingly, NGS technologies utilize reads with various larger lengths:
$k$=70 for {\sl Complete Genomics}, $35 \sim 85$ for {\sl ABI SOLiD},
$75 \sim  150$ pair-end for {\sl Illumina HiSeq}, 400 for {\sl Ion Torrent PGM}, 
$450 \sim 600$ for {\sl Roche 454 GS FLX Titanium XLR70}, etc.  \citep{ngs-rev}.
Currently, the technology is pushing towards read lengths of
$k$=1000 (e.g., {\sl Roche 454 GS FLX Titanium XL+})
or even $k$=10000 \citep{10k-news}.
Needless to say, the longer the read length, the higher the chance
that reads can be aligned to the reference genome.
Ultimately, high quality genome will be obtained by a mix of technologies.
To find this optimal mixture, a quantitative  understanding of the
repeat structure of the human genome is required.

Our analysis of the repeat structure is different from some earlier investigations
of read mappability \citep{koehler,lee}. In these studies,
the actual reads from the current sequencing technology are used.
There are two shortcomings in these approaches: (i) it is impossible
to extrapolate the result to read lengths which is beyond the
current technology; (ii) a certain proportion of reads are never mappable
because the corresponding regions in the reference genome
are not finished. Using the existing reference genome makes
it possible to treat $k$-mers as hypothetic reads whose length $k$
can be as long as possible, and unfinished regions can be excluded from the analysis.

In this paper we quantitatively address the question of how
alignment improves for greater read lengths. To this end we
artificially cut the human reference genome into overlapping
$k$-windows ($k$-mers, $k$-tuples, or $k$-gram \citep{brown}), each considered
to be possible a ``read", and count the number of appearances (or ``tokens",
borrowing a terminology from linguistics \citep{baayen}) of each $k$-mer type
across the full reference sequence.  Those $k$-mer types that appear
in the genome only once ($f$=1) are labeled singletons, and the
remainder ($f >$ 1) are non-singletons.  Intuitively, the percentage of
non-singleton reads is expected to decrease with increasing read length $k$.
Obtaining the functional form of this decay enables us  to predict
the percentage of difficult-to-align reads at longer $k$'s.

These seemingly simple calculations already encounter a ``big data" problem
on a regular-sized computer. In particular, storing counts in
a hash table requires large amount of RAM.  Suppose a $k$-mer
needs K byte to store (e.g. $K$=$k$/4), a hash table to count
all $k$-mers in the human genome would require $3K$ GByte RAM,
which quickly becomes implausible when $k$ is greater than 100.
Using a solution that is similar to other applications where
the hard disk \citep{trellis,trellis2,soap2} or computing time \cite{seqentropy}
is traded with RAM, we use a new public-domain program DSK which utilizes
the less expensive hard disk or longer CPU time to
compensate a lack of RAM \citep{dsk}. Other efficient $k$-mer
count procedures have been proposed in \citep{tallymer,jellyfish,pritchard}.

The mathematical relationship between the fraction of non-singleton
$k$-mers and $k$ predicts the fraction of putative reads that can be
mapped uniquely.  Another statistic of interest is the distribution of
$k$-mer frequencies when $k$ is fixed at a given value. This distribution
has a head and a tail, a head for low  frequency $k$-mers (including singletons),
and a tail for high frequency $k$-mers.  In the situation when these distributions
exhibit long-tails \citep{long-tail} and power-law-like trends \cite{clauset}, 
thus fitting a straight line in log-log scale, the head end is best characterized
by the frequency distribution \citep{baayen},
whereas  the tail end is better characterized by the rank-frequency
distribution commonly related to Zipf's law in quantitative linguistics
\citep{zipf}. Our analysis of these distributions provides information
on the level of redundancy in the human genome at various scales.

Locating human genome regions that cannot be uniquely mapped by
sequencing reads (which can be called ``non-uniqueome" following
the term ``uniqueome" used in \citep{koehler})
is important in any NGS-based studies. These regions may contribute the most
to the false-positive and false-negative variant callings. These may also
be hotspot for other structural variations such as indel and copy-number-variation
\citep{sharp,perry}. We will specifically examine the location of some of these redundant regions
at the $k=1000$ level.

\section*{MATERIALS AND METHODS}

\subsection*{Genome sequence data}
The human reference genome GRCh37 (hg19)
was downloaded from UCSC's genome browser
({\sl http://genome.ucsc.edu/}). The intermittent strings of N's (marking
unfinished basepairs that cannot be sequenced with the applied technology
\citep{mccarroll}) are used to partition the 22 autosomes and
2 sex chromosomes into 322 subsequences,
and $k$-mers overlapping two chromosome partitions are not allowed.

For an additional analysis on repeat-filtered sequences, strings of lowercase
letters in the reference genome  (which mark repetitive sequences identified
by the RepeatMasker program, {\sl http://www.repeatmasker.org/}) are used to partition the
genome into 3456905 subsequences with all transposable elements removed.

\subsection*{Counting $k$-mers}

A $k$-mer type includes both the direct and the reverse complement substring;
AAGC/GCTT is an example of such a 4-mer type.  We use a state-of-art $k$-mer counting
program DSK \citep{dsk} ({\sl http://minia.genouest.org/dsk/}), version 1.5031 (March 26, 2013).
Most of the DSK calculations were  carried out on a Linux computer with
48 GByte RAM and around 900 GByte disk space, except a calculation at $k$=1000
which was run on another Linux computer with the same RAM but 30 TByte of disk space.
The parameter setting of DSK  was determined by a trial-and-error process.
The output of the DSK program consists of a list of $k$-mers.  The
BLAT program from UCSC's genome browser is used to map frequent $k$-mers
back to the reference genome.

\subsection*{ Frequency distribution, rank frequency plot, and data fitting}

Suppose a $k$-mer type appears in the genome $f$ times ($f$ is frequency, or copy number);
frequency distribution (FD) is the number of $k$-mer types
with frequency $f$. Individual $k$-mer types can be ranked
by their $f$, highest $f$ ranks number 1, second highest $f$ ranks number 2,
etc. The ranked $f$'s of $k$-mer types as a function of rank $r$ is
the rank-frequency distribution (RFD).

The functions used here in fitting the RFD can all be
expressed as linear regression, include Weibull function:
$\log(f) \sim  \log( \log(( \max(r)+1)/r))$ \citep{wli-entropy};
quadratic logarithmic: $\log(f) \sim \log(r) + (\log r)^2$ \citep{wli-jql};
and reverse Beta: $\log(r) \sim \log(f) + \log( \max(f)+1-f)$. The latter
function is derived from the Beta rank function \citep{mansilla,gustavo}
by reversing the $f$ and $r$. All linear regressions are carried out
by the {\sl R} function {\sl lm} ({\sl http://www.r-project.org/}).

\section*{RESULTS}

\subsection*{Percentage of non-singleton reads vs. read length: piece-wise power-law function}

\begin{figure}[t]
 \begin{center}
 \epsfig{file=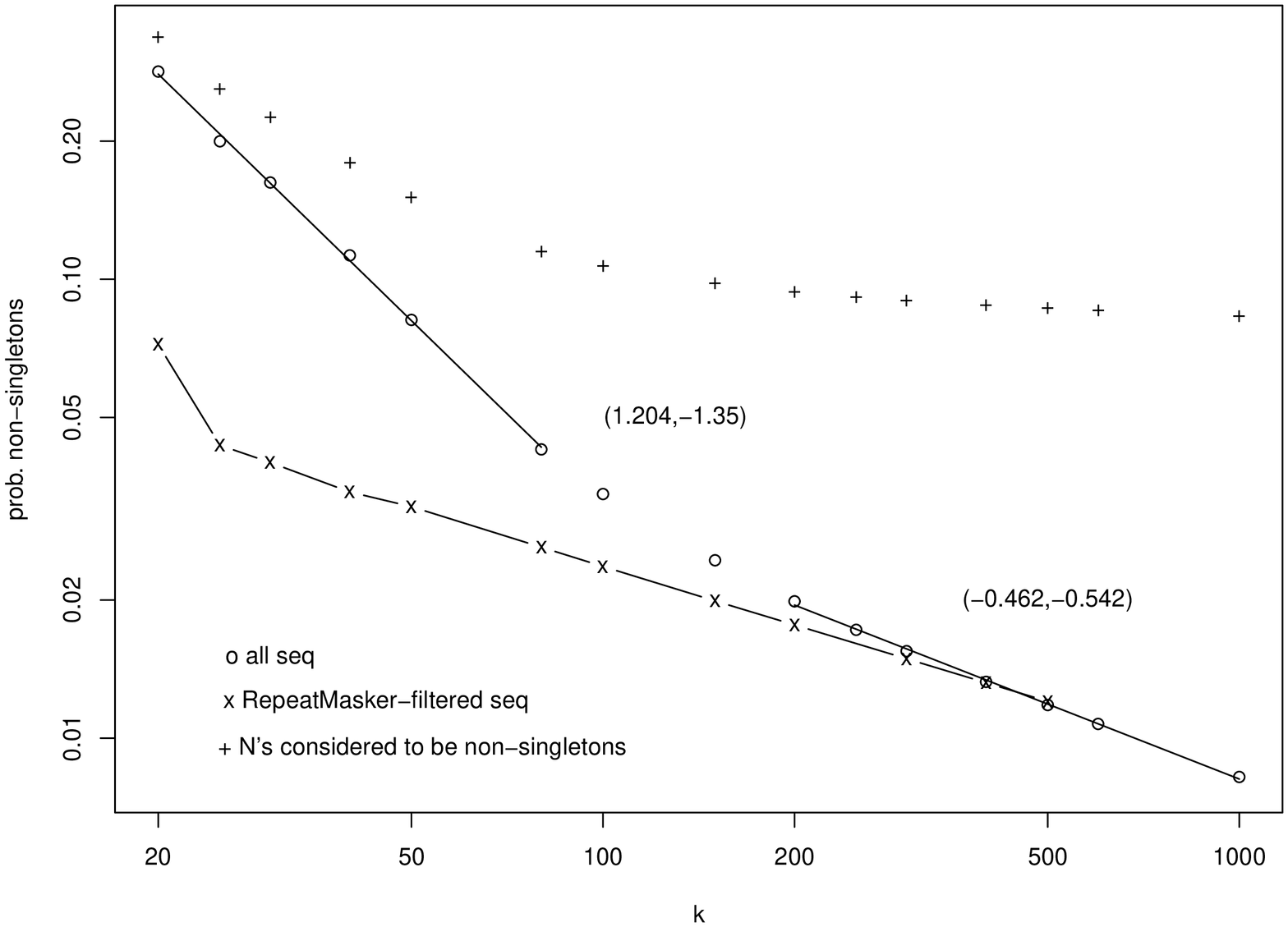, height=12cm, angle=-90}
 \end{center}
\caption{
Proportion of non-singleton $k$-mers/tokens in the human genome
(24 chromosomes) as a function
of $k$ (in log-log scale). Circles (o) show the results for all finished
basepairs, whereas crosses (x) for the result from RepeatMasker-filtered
sequences. Pluses (+) are results when unfinished sequences (234 Mbase)
are included as non-singletons.
}
\label{fig1}
\end{figure}

In Figure \ref{fig1} we  show the percentage of non-singleton reads/tokens  ($p_{ns}$)
 as a function of $k$-mer length $k$ in log-log scale.
The $p_{ns}$ is 28.35\% at $k$=20, 8.16\% at $k$=50, 4.26\% at $k$=80,
3.40\% at $k$=100, 2.44\% at $k$=150, 1.33\% at $k$=400, 1.18\% at $k$=500,
and 0.82\% at $k$=1000.  If $k$ is shorter than the ``shortest unique substring"
length, which is 11 in the human genome \citep{wiehe},
singletons do not exist (i.e., $p_{ns}=100\%$).

Visual inspection of the trend suggests the use of piecewise
power-law function in fitting the data. We fit the points in $k$ = 20-80 and
$k$= 200-1000 ranges separately by linear regressions in the log-log scale:
$\log_{10} p_{ns}= a+b \log_{10} k$ (or $\log p_{ns} \sim \log k$).
The fitted $(\hat{a}, \hat{b})$ is (1.58366, -1.5478) and (-0.4371, -0.5495)
for the two segments, equivalent to $p_{ns}= 38.34/k^{1.548}$ and $p_{ns}= 0.365/k^{0.55}$.
The steep decay in the first segment shows a stronger increase of
the amount of uniquely mappable sequences with read length, which implies
that obtaining read lengths of at least around 100 is more cost-efficient
with respect to reducing the amount of non-mappable reads.  Of course,
longer reads have extra benefits such as more robust alignments in the
presence of polymorphisms or the ability to determine the length of longer
repeat polymorphisms.  The power-law function also indicates that the
reduction of non-specific, difficult-to-align reads with longer read length is not linear.

If we assume our fitting function can be extrapolated
to larger $k$'s for which a direct analysis of $k$-mer frequencies
is restricted by computational constraints, the proportion
of non-singleton reads can be predicted.  For example, this leads
to the prediction of a 0.2\% non-singleton rate at the 10kb read length.

It is known that repetitive sequences such as transposable elements
and heterochromatin sequences create considerable obstacle in NGS alignment 
\citep{salzberg}. 
Though transposable elements may exhibit subtle correlation with functional
units in the genome \cite{wli-10mer},
it is generally assumed that their biological role is indirect.  Accordingly, we also
looked at the non-singleton $k$-mer percentages in RepeatMasker filtered sequences
(Figure \ref{fig1}). As expected, the percentage of uniquely mappable sequence
is much higher than that in the all-inclusive sequence for short $k$-mers
(e.g. $k < 100$).  Interestingly, the differences between the two disappear for
longer $k$-mers (e.g. $k$=500).  A note of caution is that 89\% of these 
RepeatMasker-filtered subsequences are shorter than 1kb, making the statistics 
less reliable at longer $k$'s.

\subsection*{Maximum $k$-mer frequency decreases with $k$ slowly}

\begin{figure}[!tpb]
 \begin{center}
  \epsfig{file=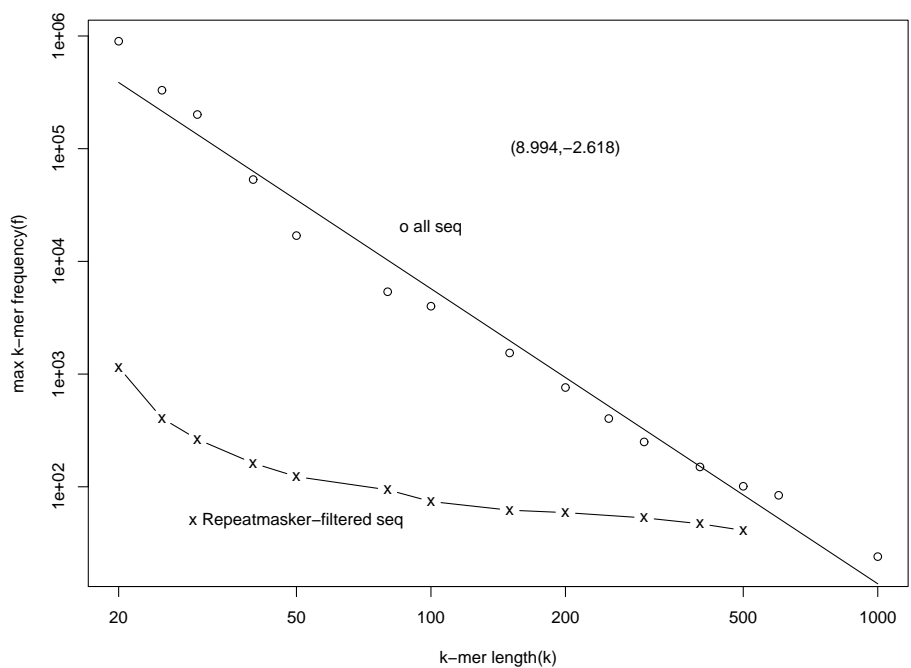, height=12cm, angle=-90}
 \end{center}
\caption{
Maximum frequencies of $k$-mers as a function of $k$
(in log-log scale). Circles (o) show the results for all finished
bases, whereas crosses (x) for the result from RepeatMasker-filtered
bases.
}
\label{fig2}
\end{figure}

Another measure of the level of redundancy at length scale $k$
is the maximum frequency (max($f$)) of $k$-mer types. For
example, base A/T homopolymers of length 20 appear most often with 898647 copies;
at $k$=400, AT repeats have more copy
numbers ($f=$150) than other 400-mers; the max($f$) for $k$=1000 is equal to 24
for a sequence which is not filtered by the RepeatMasker.
The max($f$) as a function of $k$ is shown in Figure \ref{fig2}  in log-log scale.

For RepeatMasker-filtered sequences, max($f$) quickly decays below
100 and then falls only slowly, indicating that RepeatMasker usually
finds shorter repeats. At $k$= 200-500, the $k$-mer with the
max($f$) $\sim$ 50 is a low-complexity sequence, with internal repeats
of GGGGGGAACAGCGACAC/GTGTCCGCTGTTCCCCCC. Despite its high prevalence, this 
low-complexity sequence is not masked by RepeatMasker in the human reference genome.

Fitting of the linear regression, $\log_{10} \max(f) = a +b \log_{10} k$
(or $\log \max(f) \sim \log k$),
leads to $(a,b)$= (8.99, -2.62). Extrapolating this regression to
longer $k$'s predicts that at $k$=2724, max($f$)=1. This prediction
should be viewed with caution as max($f$) is mainly determined by ``outlier" events
thus un-reproducible in principle.

\begin{figure}[!tpb]
 \begin{center}
  \epsfig{file=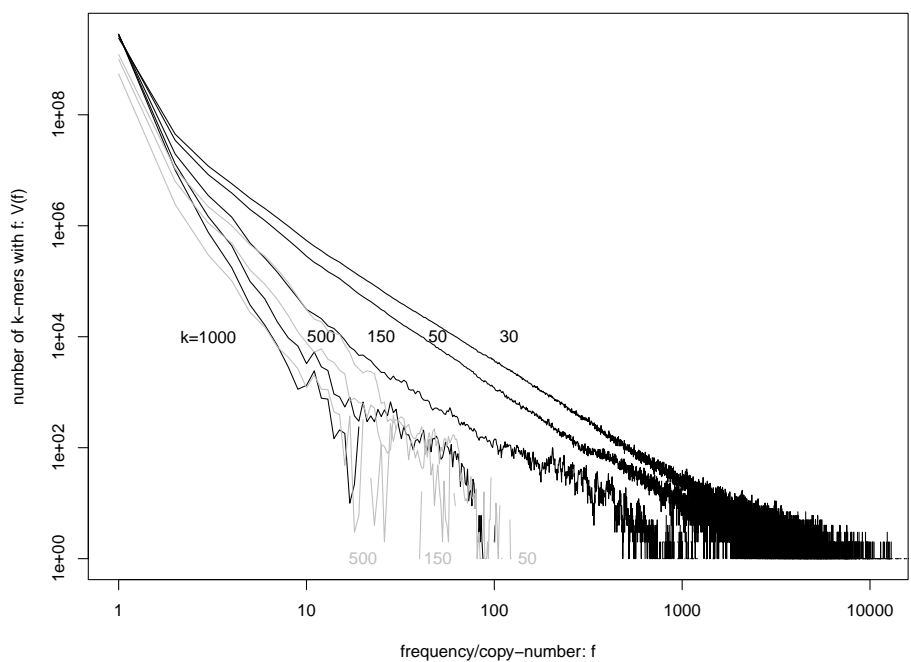, height=12cm, angle=-90}
 \end{center}
\caption{Frequency distributions of $k$-mers at $k$=30, 50, 150, 500, and 1000 (in log-log scale).
The distributions for $k$-mers in repeat-filtered sequences at $k$=50, 150, 500 are shown
in grey lines.
}
\label{fig3}
\end{figure}

{\bf Frequency distributions at fixed k values exhibit power-law-like trend:}
The frequency distribution (FD) describes the distribution of $k$-mer types
according their copy numbers in the genome. When plotted in log-log scale,
low-frequent $k$-mer types and the less redundant portion of the sequence
are highlighted. Figure \ref{fig3} shows five FDs at $k$=30, 50, 150, 500,
and 1000 in log-log scale.  The FDs at $k$=30 and 50 span a wider
frequency range, and the power-law trend is obvious.

A similar FD for $k$=40 in human genome was shown in \citep{sindi,sindi-40mer},
and a slope of $-2.3$ in linear regression (in log-log scale) in the $f$= 3-500
range was reported. When we fit the $k$=50 FD by linear regression in
log-log scale, a very similar fitting slope value is obtained ($-$2.38, for
$f$ = 3-200). However, it is clear from Figure \ref{fig3} that the slopes are
steeper for $k$=150 ($-$2.7 for $f$= 2-100), $k$=500 ($-$3.5 for $f$=2-40), and $k$=1000
($-$5.3 for $f$= 2-19, or $-$5.9 from $f$= 2-9), indicating that the slope is not a universal parameter.

From the short read alignment perspective, the long tail at the high copy-numbers
shows that many sequences cannot be uniquely mapped at smaller $k$ values
(e.g. $k$= 30, 50). However, the tail
is much shortened at  $k$=1000.  As expected, the tail for
RepeatMasker-filtered sequences at various $k$ values are much
shorter (Figure \ref{fig3}, grey lines).

\subsection*{Rank-frequency distributions at fixed k values mostly follows a
concave curve in log-log scale}

\begin{figure}[!tpb]
 \begin{center}
  \epsfig{file=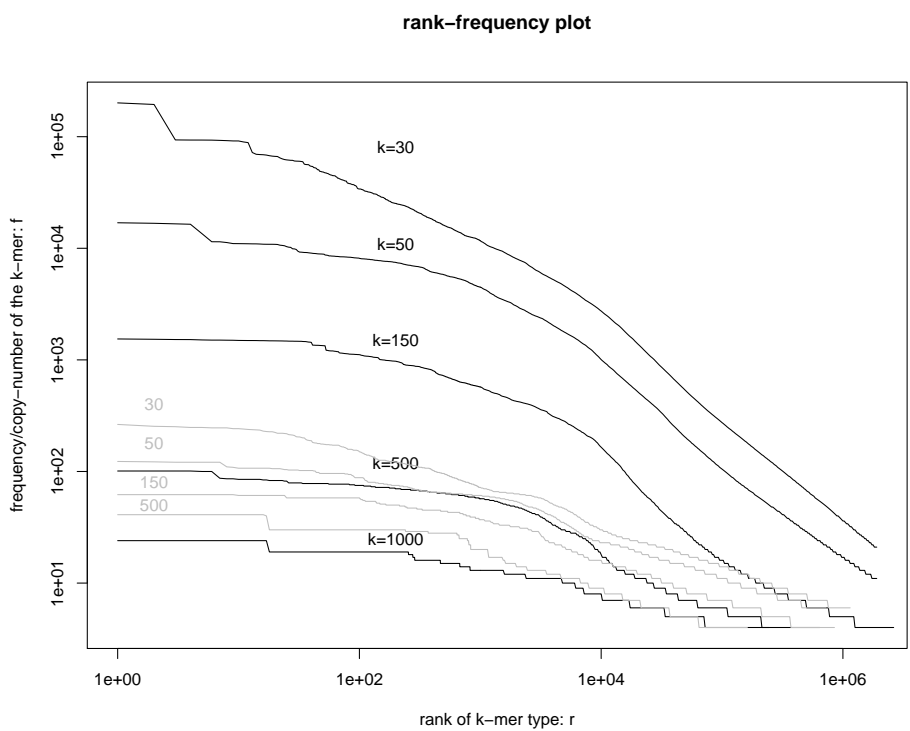, height=12cm, angle=-90}
 \end{center}
\caption{Rank-frequency distributions for $k$-mers at $k$= 30, 50, 150, 500, and 1000
(in log-log scale). The corresponding rank-frequency distributions for RepeatMasker-filtered
sequences at $k$= 30, 50, 150, 500 are shown in grey lines.
}
\label{fig4}
\end{figure}

Although rank-frequency distributions (RFD's) can be converted
to cumulative FD \citep{wli-entropy}, in log-log scale, it zooms in
the high-frequency tail of the frequency distribution.  Figure \ref{fig4}
shows five RFD at $k$'s from 30 to 1000.  While the RFD at $k$=30 may maintain a power-law
or piecewise power-law trend, those at larger $k$ values become more concave.
This concave Zipf's curve is commonly observed in city size
distributions \citep{gabaix04,gibrat}.

\begin{figure}[!tpb]
 \begin{center}
  \epsfig{file=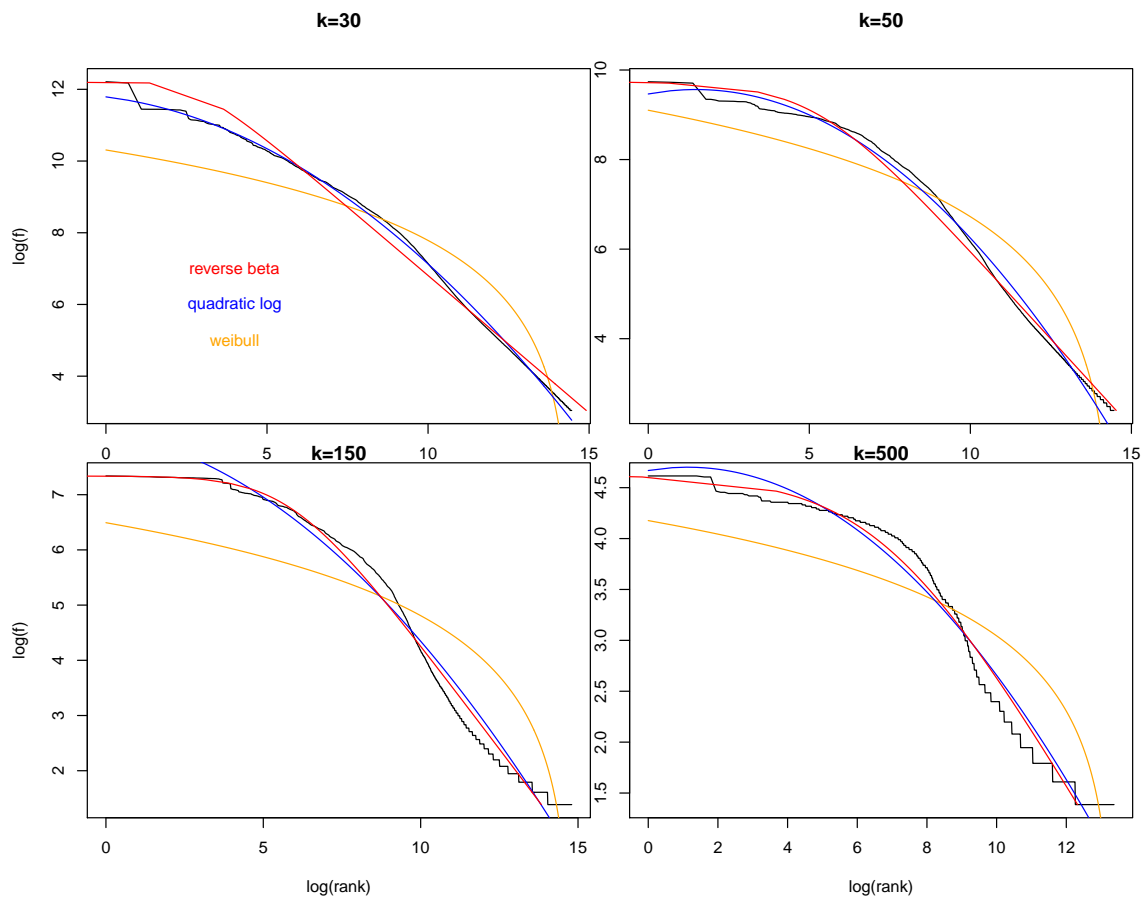, height=15cm, angle=-90}
 \end{center}
\caption{
Fitting rank-frequency distribution of $k$-mers at $k$=30, 50, 150, 500 using three functions.
Red: quadratic logarithmic ($\log f \sim \log(r) + \log((r))^2$, $f$: frequency of
a $k$-mer type, $r$: rank of a $k$-mer type, and the $\sim$ symbol represents linear regression); 
blue: reverse Beta rank function ($\log(r) \sim \log(f) + \log( max(f)+1-f) $); 
Orange: Weibull function ($\log(f) \sim  \log( \log( (max(r)+1)/r) )$). 
}
\label{fig5}
\end{figure}

For RFDs deviating from the Zipf's law, functions with two parameters
may be used to account for the concave or convex shape of the curve
in log-log scale \citep{wli-entropy}. We found that the quadratic
logarithmic function, but not the Weibull function, fits the RFDs well 
(Figure \ref{fig5}).
The Beta rank function usually exhibit ``S" shapes
\citep{gustavo}, whereas the RFD in Figure \ref{fig4}
shows a ``Z" shape. This motivated us to use a novel
reverse Beta function to fit the data (Figure \ref{fig5}).
The ``Z" shaped log-log RFD means that if the power-law function
is the default functional relationship between frequency and rank,
frequencies of the intermediately-ranked $k$-mers decrease faster
than the two tails. The ``S" shaped log-log RFD implies the opposite.

\subsection*{Mapping $f \ge 10$ 1000-mer to the reference genome}

For $k$=1000, there
are 6107 $k$-mer types with frequency $f$ larger or equal to 10. Due to the
fact that these are overlapping $k$-mers, they are mapped to only
172  chromosomal regions, each of a few kb 
(the 172 locations, number of high-frequency 1000-mers, and the distance 
from the previous chromosome regions are included in Supplementary Table S1).

A total of 70 out of these 172 regions (or 40\%) are clustered
in four larger stretches on chromosomes 1 and X and contain long
tandem repeats (60, 70 kbase on chromosome 1q21.1, 1q21.2, and  41, 56 kbases
on Xq23, Xq24).  The two stretches on chromosome 1
contain copies of the neuroblastoma breakpoint family genes
(NBPF) \citep{nbpf,nbpf-chimp,nbpf-human}. The Xq24 region
contains cancer/testis antigen family genes (CT47A) \citep{ct47a,dobrynin},
whereas the Xq23 region has no genes, but contains the macrosatellite
DXZ4 \citep{gia,dxz4,macro} which exhibits periodic appearance of other functional
elements, such as H3K27Ac or H3K4me2 \citep{hora} histone modification marks.

Besides these long stretches, 39 out of 172 regions (or 23\%)
overlap with 34 genes: {\sl ZNF3850, EPHA3, COL6A6,  CD38, KCNIP4, FRAS1, ANTXR2,
HSD17B11, FAM190A, DKK2, FBXL7, AK123816, FAM153A, FAM65B,
LAMA2, MYCT1, NOD1, TPST1, PSD3, KCNB2, NR4A3, C9orf171,
CACNA1B, DLG2, CCDC67, UACA, HOMER2, SMG1, CDH13 ,
PRKCA, LILRA2, TTC28, MTMR8}, and {\sl SLC25A43}.
Obtaining high quality data on genetic variants in these gene sequences
is therefore likely to remain a challenge even with longer reads.

\section*{DISCUSSION}

{\bf Long $k$-mers in the reference genome as surrogate for sequencing reads:}
The $k$-mer distribution has many application in sequence analysis, such
as measuring similarity between two genomes \citep{edgar}, correcting 
sequencing error \citep{musket}, finding repeat structures \cite{waterman}, 
determining the feasibility of gene patents \citep{patent}.
In many applications, only short $k$-mers are considered to be relevant, such as 
$k=6$ \citep{chen}, $k \le 7$ \cite{chris}, $k$=8 \cite{hao},
$k$=11 \citep{chor}. This paper essentially uses long $k$-mers
taken from the reference genome as surrogate for reads from
future NGS technologies.  Computationally speaking, counting 
long $k$-mers is more challenging and we are not aware of any prior 
publications on the long $k$-mer distributions in the human genome
for $k$ as long as 1000.

As compared to other papers on mappability of genome sequencing reads 
\citep{koehler,lee}, our more theoretical approach has the advantage of being
able to discuss long reads (e.g. $k$=1000) where such data is
not available from the current NGS technology. Our approach also
separates the two causes of the unmappability: one due to the 
unfinished sequence in the reference genome and another due to
the redundancy in the finished sequences. 
The unfinished bases are mainly located in the centromeres, short arms
of acrocentric chromosomes and other heterochromatic regions, and rich
in repetitive sequences.  If we always treat this unfinished 
sequences (total 234 Mbases) to be non-singletons regardless 
of $k$, $p_{ns}$ would flatten out around 0.1 (see Figure \ref{fig1}).

{\bf A baseline knowledge of redundancy of the human genome at length $k$ level:}
Figures 1-3 provides a baseline knowledge of the redundancy of
the human genome at the $k$-mer level.  Our results give a more 
quantitative description on the effect of read length $k$ on 
the mappability of reads from the finished region of the human genome.

Reference assembly is 
easier than {\sl de novo}
assembly, and our approach does not directly apply to {\sl de novo} 
sequencing ``assemblability". However the mappability in reference
assembly and assemblability in {\sl de novo} assembly are closely
related, as repetitive sequences cause problems in both situations
\citep{pas}.
The current {\sl de novo} assemblies still do not perform
consistently \citep{ass2,clay} and a quantitative assessment of the impact 
of repetitive sequences on reference assembly could be a useful 
piece of information for {\sl de novo} assembly as well.
Note that some discussion on $k$-mer-based assembly actually refers
to $k'$-mer ($k' << k$) 
\citep{birney,liu}.

{\bf Highly redundant regions at $k=1000$ level and copy-number-variation regions:}
The chromosome 1 and X regions which we have identified by showing at least 10
copy numbers of 1000-mers are discussed in the literature as
regions with common copy-number-variations (CNV). CNV in the 1q21.1 region,
if not NBPF-specific, has been linked to congenital cardiac defect 
\citep{christian,redon,greenway}, 
autism \citep{szatmari,giri}, mental retardation \cite{mefford}, 
developmental  abnormalities \citep{brunetti}, schizophrenia \cite{schiz,ikeda}, 
and neuroblastoma \citep{nbpf-cnv}. With so many abnormalities mapped to
this region, these are collectively called the chromosome 1q21.1
duplication syndrome in the Online Mendelian Inheritance in Man (OMIM 612475). 

The Xq23 region, if not macrosatellite DXZ4 specific,  has been identified as 
likely CNV regions linked to developmental and behavioral problems \citep{isrie}.
Chromatin configuration at DXZ4 region is reported to differ between male
melanoma cells and normal skin cells \citep{moseley}. The Xq24 region
if not the CT47A gene is listed as a candidate CNV region for
intellectual disability \citep{whibley}, mental retardation \cite{honda}, etc. 

A well-known mechanism for CNV formation is the non-allelic homologous
recombinations (NAHR) between repetitive elements \citep{gu}. More
copies of a repetitive sequence give more opportunities that NAHR
could occur, resulting in a natural connection between repetitive
sequences and common CNV.   The fact that simple counting of 
1000-mer frequencies leads to CNV regions with medical 
implications indicates that understanding the  $k$-mer distribution is
an important part of genomic analyses.

{\bf Long-tails and the diminishing return of longer reads:} 
Our analysis shows that all distributions discussed in this paper are better viewed in
log-log scale, proving the existence of power-law distributions or long-tails. 
This has been observed in the past for other genomic distributions, 
such as correlation function \citep{wli-epl,wli-review,pedro,arneodo}, power spectrum of base composition
\citep{voss,fukushima,wli-dirk2,wli-dirk}, frequency distribution  of gene or protein 
family size \citep{erik,qian,koonin,herrada}, sizes of 
ultraconserved regions \citep{miller}, and in models with duplications 
\citep{wli-pra,delisi,teich,arndt}.  Ongoing duplications increase the copy number 
geometrically, which explains the presence of long-tails. 

A consequence of the long-tail in Figure \ref{fig1} is that with
increasing read (or $k$-mer) lengths, the proportion of reads that
cannot be mapped to a unique genomic region (within the finished sequences) 
decreases algebraically, as compared to linearly or exponentially.
Numerically, if not economically,  this defines a diminishing return.
To assess the economic return with NGS technology with longer reads,
other factors should be considered, such as the choice of less 
redundant target regions such as exome \citep{ng}, the choice of pair-end 
sequencing technologies \citep{pairend} which effectively increases the read length
\citep{rama} though the mappability may not necessarily improve over the
single-end sequencing \citep{derrien}, 
and the overall cost of longer-read sequencing. Anticipating the eventual
high-quality human genome sequences obtained by a combination of various
technologies, the $k$-mer distribution will be a prominent factor
determining how these technologies are optimally combined.

\section*{SUPPLEMENTARY DATA}

Supplementary Data are available at NAR online, including Supplementary
Table S1: 172 chromosome locations with high-frequency ($f \ge 10$) 1000-mers.

\section*{ACKNOWLEDGEMENTS}

We would like to thank Oliver Clay, Andrew Shih, Astero Provata, Yannis Almirantis for
discussions, and the authors of DSK for timely responding to our inquiries and fixing bugs.
WL and JF acknowledge the support from the Robert S Boas Center for Genomics and Human Genetics.

\newpage

\clearpage

\clearpage

\clearpage

\clearpage

\newpage

\normalsize

\section*{ Supplementary material: Table S1}

\indent

Chromosome locations of $k$=1000-mers with frequency $f \ge 10$ mapped to the human reference
genome (GRCh37/hg19, Feb 2009). The seven columns are:
1. chromosome (23 for chromosome X);
2. starting position (in base);
3. ending position (in base);
4. width of the region (in kbases);
5. number of 1000-mers mapped to this region;
6. spacing with the previous region (in base);
7 (if available) gene name.

\small

\begin{verbatim}
ch      start   end     width(kb)   num-entity dist-prev-end gene-name
 1  56831364  56833043   1.679  390 56831364 
 1  84518510  84519720   1.210  162 27685467 
 1  84521988  84523534   1.546  333     2268 
 1 144163689 144164940   1.251  253 59640155   NBPF-family
 1 144164949 144166172   1.223  156        9   NBPF-family
 1 144166529 144168209   1.680  682      357   NBPF-family
 1 144171267 144172868   1.601  603     3058   NBPF-family
 1 144174458 144177621   3.163 1238     1590   NBPF-family
 1 144178038 144179323   1.285  287      417   NBPF-family
 1 144179746 144180804   1.058   60      423   NBPF-family
 1 144180830 144184012   3.182  710       26   NBPF-family
 1 144190348 144193522   3.174  961     6336   NBPF-family
 1 144193538 144195088   1.550  552       16   NBPF-family
 1 144195507 144201554   6.047 1426      419   NBPF-family
 1 144203065 144204372   1.307  264     1511   NBPF-family
 1 144206755 144207807   1.052   54     2383   NBPF-family
 1 144207827 144209509   1.682  684       20   NBPF-family
 1 144211012 144214175   3.163 1238     1503   NBPF-family
 1 144214195 144215877   1.682  684       20   NBPF-family
 1 144217378 144220552   3.174  961     1501   NBPF-family
 1 144222542 144223699   1.157  154     1990   NBPF-family
 1 148255552 148256803   1.251  253  4031853   NBPF-family
 1 148257416 148258725   1.309  260      613   NBPF-family
 1 148260316 148261567   1.251  253     1591   NBPF-family
 1 148263509 148265059   1.550  552     1942   NBPF-family
 1 148265073 148268247   3.174  710       14   NBPF-family
 1 148268281 148269831   1.550  552       34   NBPF-family
 1 148271714 148273021   1.307  264     1883   NBPF-family
 1 148273051 148277793   4.742 1139       30   NBPF-family
 1 148278120 148280976   2.856  931      327   NBPF-family
 1 148282487 148284169   1.682  684     1511   NBPF-family
 1 148285680 148286965   1.285  287     1511   NBPF-family
 1 148287392 148288944   1.552  554      427   NBPF-family
 1 148288961 148291735   2.774  563       17   NBPF-family
 1 148292162 148293714   1.552  554      427   NBPF-family
 1 148293731 148296505   2.774  563       17   NBPF-family
 1 148296932 148298484   1.552  554      427   NBPF-family
 1 148298501 148301275   2.774  561       17   NBPF-family
 1 148301694 148303244   1.550  552      419   NBPF-family
 1 148303258 148306035   2.777  311       14   NBPF-family
 1 148308035 148309286   1.251  253     2000   NBPF-family
 1 148311228 148312780   1.552  554     1942   NBPF-family
 1 148312794 148315968   3.174  963       14   NBPF-family
 1 148317640 148318820   1.180  101     1672   NBPF-family
 1 148320748 148322300   1.552  554     1928   NBPF-family
 1 148322314 148325488   3.174  708       14   NBPF-family
 1 237186076 237187264   1.188   15 88860588 
 1 247852722 247854662   1.940  370 10665458 
 2   4782957   4784039   1.082   68       NA
 2  11138506  11139623   1.117   51  6354467 
 2 134968954 134970563   1.609  162 123829331 
 3  22092360  22093571   1.211   63       NA      ZNF3850
 3  22096119  22097785   1.666  577     2548       ZNF3850
 3  80925499  80926606   1.107   34 58827714 
 3  89513287  89515334   2.047  638  8586681         EPHA3
 3 108920256 108921439   1.183   93 19404922 
 3 130351657 130352915   1.258  260 21430218        COL6A6
 3 137072367 137073449   1.082   84  6719452 
 4  15845170  15846306   1.136  138       NA         CD38
 4  21163013  21164348   1.335  237  5316707        KCNIP4
 4  21165563  21167050   1.487  448     1215        KCNIP4
 4  71194736  71195804   1.068   40 50027686 
 4  75644610  75646550   1.940  370  4448806 
 4  79271909  79273254   1.345   73  3625359         FRAS1
 4  80863626  80864733   1.107   34  1590372        ANTXR2
 4  80890063  80891544   1.481  131    25330 
 4  88268648  88269936   1.288  210  7377104      HSD17B11
 4  88270964  88272227   1.263  165     1028      HSD17B11
 4  88272229  88273386   1.157  159        2      HSD17B11
 4  91600685  91602074   1.389  147  3327299       FAM190A
 4 108005014 108277809 272.795  128 16402940          DKK2
 4 137214667 137216333   1.666  577 28936858 
 4 137217568 137220031   2.463  776     1235 
 4 139471548 139473214   1.666  577  2251517 
 5     10479     11579   1.100   60       NA
 5  15906794  15907818   1.024   26 15895215         FBXL7
 5  57682240  57683559   1.319   49 41774422 
 5 103855313 103857622   2.309  547 46171754 
 5 108595761 108598576   2.815  772  4738139 
 5 152268410 152269605   1.195  120 43669834      AK123816
 5 166395979 166397375   1.396  381 14126374 
 5 177201688 177203367   1.679  390 10804313       FAM153A
 5 177203999 177205005   1.006    8      632       FAM153A
 6  24814196  24815833   1.637  335       NA
 6  24816124  24817135   1.011   13      291 
 6  86712093  86713844   1.751  383 61894958 
 6 129321768 129322865   1.097   99 42607924         LAMA2
 6 133342240 133343397   1.157  141  4019375 
 6 133344551 133346228   1.677  285     1154 
 6 153032481 153033556   1.075   77 19686253         MYCT1
 7  30481972  30482979   1.007    9       NA         NOD1
 7  32390477  32392004   1.527  176  1907498 
 7  49720296  49721702   1.406    7 17328292 
 7  49723677  49725225   1.548  525     1975 
 7  65756060  65757315   1.255   41 16030835         TPST1
 7  96479274  96481321   2.047  638 30721959 
 7 113416172 113417537   1.365  367 16934851 
 7 141620623 141622142   1.519  419 28203086 
 8  73787961  73789068   1.107   34 55332129         KCNB2
 8 126597079 126598414   1.335  237 52808011 
 8 126599450 126600465   1.015   17     1036 
 8 129468131 129469267   1.136  138  2867666 
 8 129469381 129470594   1.213  215      114 
 8 135086298 135087380   1.082   68  5615704 
 8  18454681  18455832   1.151  153       NA         PSD3
 9  96880037  96881079   1.042   44       NA
 9  98464614  98465721   1.107   34  1583535 
 9 102615779 102616803   1.024   26  4150058         NR4A3
 9 135401989 135402996   1.007    9 32785186      C9orf171
 9 140912461 140914051   1.590  592  5509465       CACNA1B
10 107137274 107138381   1.107   34       NA
10 111574095 111575101   1.006    8  4435714 
11  24351213  24352216   1.003    5       NA
11  24354168  24355533   1.365  367     1952 
11  60852454  60853848   1.394  373 36496921 
11  85038244  85039346   1.102  104 24184396          DLG2
11  93155588  93156702   1.114   42  8116242        CCDC67
11  93158897  93160004   1.107   34     2195        CCDC67
11  95170052  95171840   1.788  545  2010048 
11  95173750  95175002   1.252  204     1910 
11 125410876 125411942   1.066   68 30235874 
12  75271094  75272762   1.668  318       NA
12  88141507  88142614   1.107   34 12868745 
12 126784020 126785217   1.197  149 38641406 
12 126787028 126788915   1.887  570     1811 
13  30220115  30221202   1.087   89       NA
14  63587645  63588655   1.010   12       NA
15  55220178  55221857   1.679  390       NA
15  71023909  71025588   1.679  390 15802052          UACA
15  83555788  83556968   1.180  182 12530200        HOMER2
16    236025    237244   1.219  200       NA
16  16936305  16937984   1.679  390 16699061 
16  18834487  18836369   1.882  489  1896503          SMG1
16  83671524  83673049   1.525  174 64835155         CDH13
17  64594609  64595616   1.007    9       NA        PRKCA
17  68457097  68458432   1.335  237  3861481 
17  68459468  68461134   1.666  577     1036 
18  68415980  68417774   1.794  319       NA
19  55091938  55092944   1.006    8       NA       LILRA2
22  29064027  29065134   1.107   34       NA        TTC28
23  11730147  11731404   1.257  259       NA
23  11957210  11958761   1.551  528   225806 
23  63474347  63475805   1.458   33 51515586         MTMR8
23  81101312  81102523   1.211  202 17625507 
23 114959682 114961907   2.225 1227 33857159 macrosatellite-DXZ4
23 114962577 114964888   2.311 1313      670 macrosatellite-DXZ4
23 114965558 114967514   1.956  958      670 macrosatellite-DXZ4
23 114968541 114969796   1.255  257     1027 macrosatellite-DXZ4
23 114969798 114970852   1.054   56        2 macrosatellite-DXZ4
23 114971520 114973552   2.032  871      668 macrosatellite-DXZ4
23 114974777 114976822   2.045 1047     1225 macrosatellite-DXZ4
23 114977502 114979813   2.311 1313      680 macrosatellite-DXZ4
23 114980485 114982796   2.311 1313      672 macrosatellite-DXZ4
23 114983470 114985781   2.311 1313      674 macrosatellite-DXZ4
23 114986449 114988760   2.311 1313      668 macrosatellite-DXZ4
23 114989702 114991747   2.045 1047      942 macrosatellite-DXZ4
23 114992405 114994716   2.311 1313      658 macrosatellite-DXZ4
23 114995398 114997709   2.311 1313      682 macrosatellite-DXZ4
23 114998885 115000439   1.554  224     1176 macrosatellite-DXZ4
23 118572124 118573459   1.335  237  3571685      SLC25A43
23 120064672 120066902   2.230 1232  1491213  CT47A-family
23 120069533 120071763   2.230 1232     2631  CT47A-family
23 120074394 120076624   2.230 1232     2631  CT47A-family
23 120079254 120081484   2.230 1232     2630  CT47A-family
23 120084115 120086345   2.230 1232     2631  CT47A-family
23 120088975 120090501   1.526  351     2630  CT47A-family
23 120094352 120096088   1.736  707     3851  CT47A-family
23 120098719 120100949   2.230 1232     2631  CT47A-family
23 120103579 120105809   2.230 1232     2630  CT47A-family
23 120108439 120110669   2.230 1232     2630  CT47A-family
23 120113299 120115529   2.230 1232     2630  CT47A-family
23 120118159 120120389   2.230 1232     2630  CT47A-family
\end{verbatim}

\end{document}